\documentclass[manuscript]{acmart}

\usepackage[shortlabels]{enumitem}
\usepackage{CJKutf8}
\usepackage{xcolor}
\usepackage{hyperref}
\usepackage[normalem]{ulem}
\usepackage{enumitem}
\usepackage{caption}
\usepackage{subcaption}
\usepackage{wrapfig}

\AtBeginDocument{%
  \providecommand\BibTeX{{%
    \normalfont B\kern-0.5em{\scshape i\kern-0.25em b}\kern-0.8em\TeX}}}

\begin{document}

\title{Recommendations and Results Organization in Netflix Search}

\author{Sudarshan Lamkhede}
\email{slamkhede@netflix.com}
\orcid{0000-0001-8699-3776}
\affiliation{%
  \institution{Netflix Inc.}
  \streetaddress{100 Winchester Boulevard}
  \city{Los Gatos}
  \state{California}
  \postcode{95032}
  \country{United States}
}
\author{Christoph Kofler}
\email{ckofler@netflix.com}
\affiliation{%
  \institution{Netflix Inc.}
  \streetaddress{100 Winchester Boulevard}
  \city{Los Gatos}
  \state{California}
  \postcode{95032}
  \country{United States}
}

\newcommand{\search}{Search}
\newcommand{\instantsearch}{Instant Search}
\newcommand{\recsys}{Recommender Systems}
\newcommand{\userquery}[1]{\texttt{#1}}
\newcommand{\movietv}[2]{\href{#2}{#1}}
\newcommand{\suggest}[1]{{\color{red} #1}}

\renewcommand{\shortauthors}{Lamkhede and Kofler}


%

\maketitle

\section{Introduction}
Personalized \textit{recommendations} on the \textit{Netflix Homepage} are based on a user's viewing habits and the behavior of similar users. These recommendations, organized for efficient browsing, enable users to discover the next great video to watch and enjoy without additional input or an explicit expression of their \textit{intents} or goals.
The \textit{Netflix Search} experience, on the other hand, allows users to take active control of discovering new videos by explicitly expressing their entertainment needs via search queries. 

In this talk, we discuss the importance of producing search results that go beyond traditional keyword-matches to effectively satisfy users' search needs in the Netflix entertainment setting. Motivated by users' various search intents, we
highlight the necessity to improve Search by applying approaches that have historically powered the Homepage. 
Specifically, we discuss our approach to \textit{leverage recommendations in the context of Search} and to effectively \textit{organize search results} to provide a product experience that meaningfully adds value for our users.

\section{Search Intents and Query Facets}
Similar to other (video) search engines, when users search on Netflix they have a particular \textit{intent}, i.e., an immediate reason, purpose or goal in mind \cite{kofler_2016_intent}. 
From qualitative and quantitative data we observe that search intents fall on a spectrum between \textit{Fetching} a specific video from the catalog (\textit{"I know what I want, I need you to get it for me"}) to extensively \textit{Exploring} the catalog (\textit{"I don't know what I want, let's understand what you have"}) \cite{lamkhede_sigir_2019}. 

\begin{figure}[htbp]
   \centering
    \includegraphics[width=0.75\linewidth]{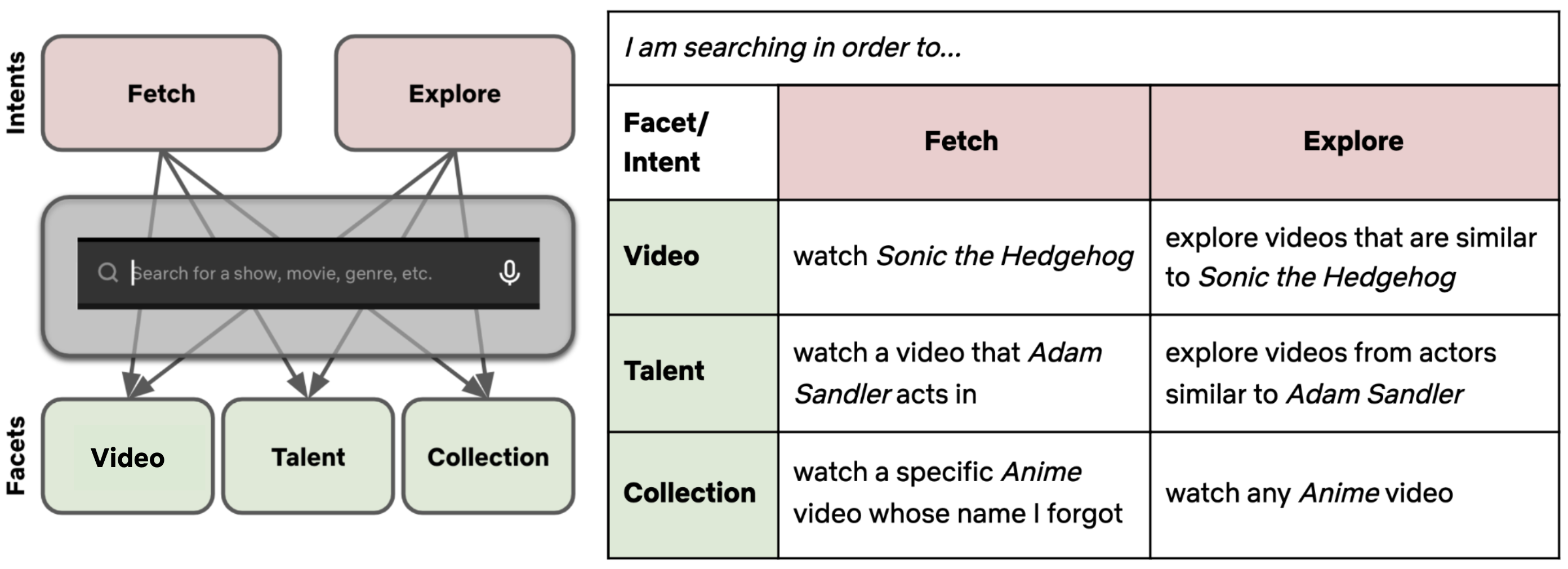}
    \caption{Illustration of the relationship between \textit{search intents} and \textit{query facets}, along with representative examples.}
    \label{fig:intents-facets}
\end{figure}

We also observe that users express their intents using different \textit{query facets}: \textit{(available and unavailable) videos} to stream on Netflix\footnote{Due to licensing and other constraints, not every video is available to stream on Netflix or comparable entertainment platforms.}, \textit{talent} (e.g., actors), and \textit{collections} (e.g,. genres). 
Fig. \ref{fig:intents-facets} illustrates the difference between intents and facets; e.g., a user searching for a specific video (i.e., the query facet), may have an intent to either play that video (\textit{Fetch}) or to explore content that is similar to that video (\textit{Explore}). By understanding the query facet, we can optimize for both intents.

\section{Motivation}

To address search intents and provide a \textit{meaningful} product experience for users, we need to rethink how \textit{effective} search results should be defined in the entertainment context. We define a \textit{search match} as a video retrieved by the search engine by \textit{keyword-matching} the query with the indexed videos (or by applying techniques such as query expansion \cite{imani_query_expansion_2018}). A \textit{search recommendation}, on the other hand, is a video selected by the search engine by relaxing the match constraints, i.e., a video retrieved via traditional recommender systems approaches (e.g., collaborative filtering) \textit{in the query context}. We use the term \textit{search results} to refer to the union of search matches and search recommendations, i.e., all videos returned in response to a user query. 

\begin{figure}
     \centering
     \begin{subfigure}[b]{0.24\textwidth}
         \centering
         \includegraphics[width=\textwidth]{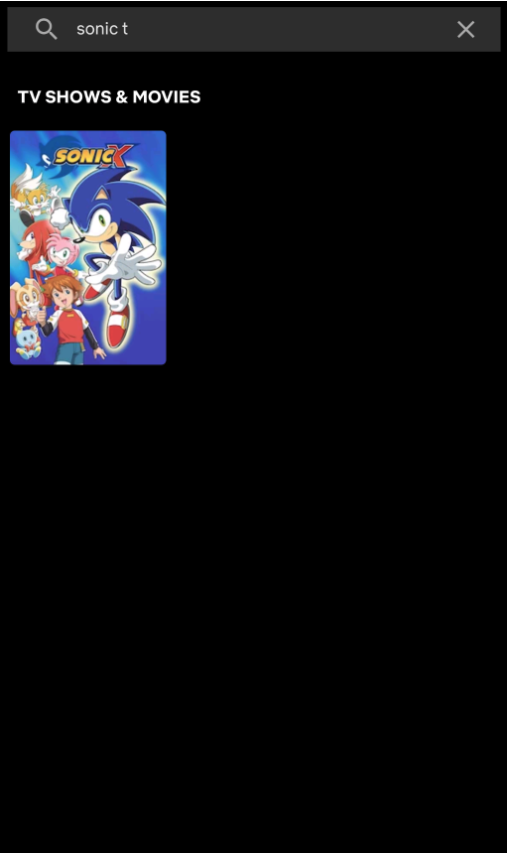}
         \caption{Traditional exact query-video keyword search matches only}
         \label{fig:1-lexical}
     \end{subfigure}
     \hfill
     \begin{subfigure}[b]{0.24\textwidth}
         \centering
         \includegraphics[width=\textwidth]{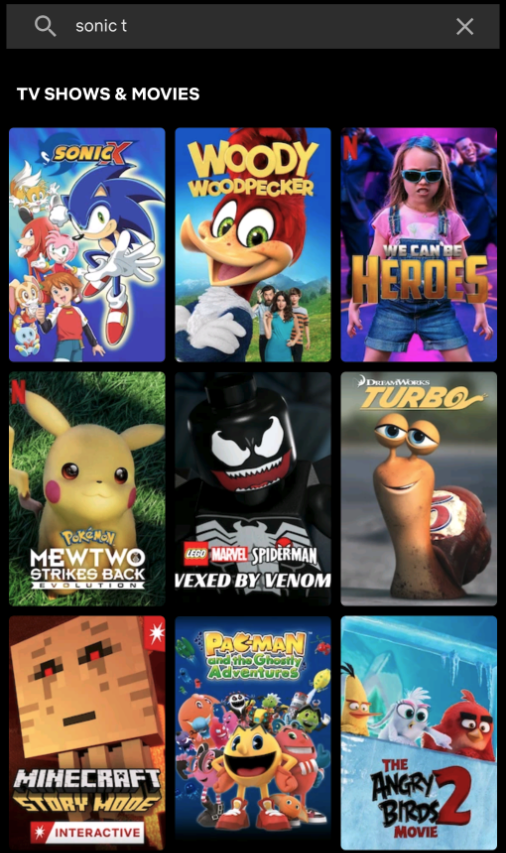}
         \caption{Recommendations relevant in the search query context}
         \label{fig:2-recs}
     \end{subfigure}
     \hfill
     \begin{subfigure}[b]{0.24\textwidth}
         \centering
         \includegraphics[width=\textwidth]{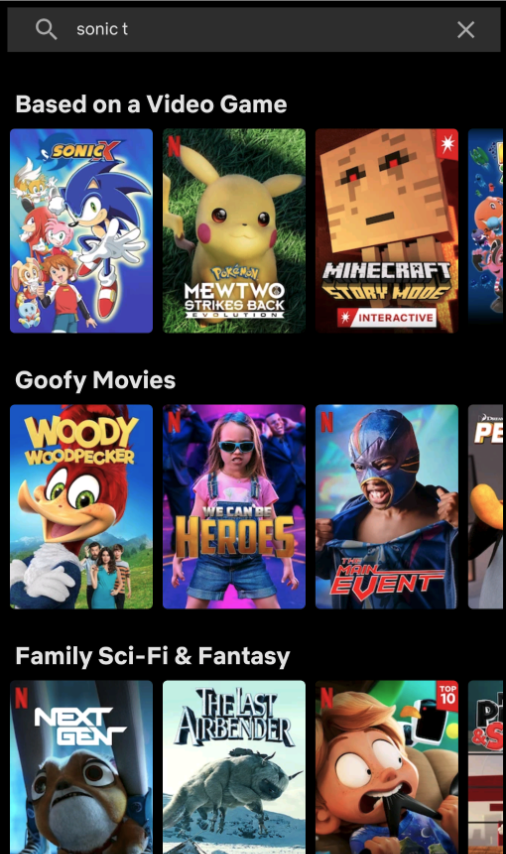}
         \caption{Results organization helping explainability}
         \label{fig:3-lolomo}
     \end{subfigure}
     \hfill
     \begin{subfigure}[b]{0.24\textwidth}
         \centering
         \includegraphics[width=\textwidth]{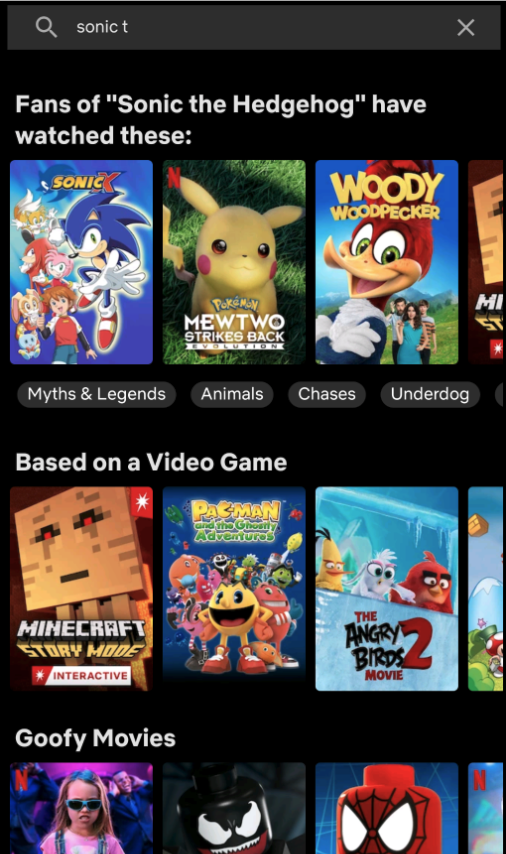}
         \caption{Results organization for \textit{unavailable video} searches}
         \label{fig:4-ooc-lolomo}
     \end{subfigure}
     
     \caption{Evolution of \textit{search results} and their \textit{organization} provided in the query context.}
     \label{fig:search-recs-evolution}
\end{figure}

\textbf{Why do we need to provide search recommendations?}
Using a traditional information retrieval approach that solely produces search matches, we would only be able to address (a subset of) \textit{Fetch} intents, and entirely neglect \textit{Explore} intents. As shown in Fig. \ref{fig:1-lexical}, when a user searches for \userquery{"sonic t"} with an intent to watch \movietv{\textit{Sonic the Hedgehog}}{https://www.imdb.com/title/tt3794354/}, we would only return one search result (i.e., \movietv{\textit{Sonic X}}{https://www.imdb.com/title/tt0367413/}) from our catalog that matches the typed in query. By producing search recommendations, however, we can uncover videos that users most associate with the query (and the query facet) and highlight aspects of our catalog that also satisfy \textit{Explore} intents (cf. Fig. \ref{fig:2-recs}). Search matches and search recommendations are roughly equally important for our users to satisfy their intents. 

Search recommendations can offer meaningful value beyond satisfying search intents by reducing friction in various ways. First, they allow us to produce effective responses to queries that target \textit{unavailable} videos. In our example, the video \movietv{\textit{Sonic the Hedgehog}}{https://www.imdb.com/title/tt3794354/} is in fact unavailable (as depicted, only \movietv{\textit{Sonic X}}{https://www.imdb.com/title/tt0367413/} is available), but we can still produce recommendations for similar \textit{available} videos that are relevant to the query \userquery{"sonic t"}, and thus help avoid "dead ends" that users may otherwise experience. Second, sometimes users do not know what to search for to fetch a video (e.g., when they forgot the name but remember other details \cite{tip_of_tongue_chiir_2021}) or there may not be an obvious query for what they are looking for (e.g., videos that are similar to the ones \textit{Adam Sandler} acts in). Recommendations reduce the need to reformulate queries in such cases. Third, when users  have an \textit{Explore} intent, recommendations may help them generalize or serendipitously pivot to related content (e.g., users may become interested in \textit{Jackie Chan} movies when they are searching for \textit{Bruce Lee}). Lastly, even when users with \textit{Fetch} intents are searching for a video that is available, knowing there are more relevant videos available provides assurance about the depth of the catalog.

\textbf{Why do we need to organize search results?} When the results produced in response to a query go beyond obvious search matches, we need to explain how they are relevant to the query. In addition, we need to reduce the cognitive burden on users to help them efficiently determine which videos satisfy their entertainment needs. 
We can achieve this by, e.g., organizing results into cohesive groups of videos that are relevant to the query (cf. \cite{hao_dumais_2000}). In our example (cf. Fig. \ref{fig:3-lolomo}), we observe that each group of videos is relevant to the query from different perspectives (e.g., \textit{Based on a Video Game}, \textit{Goofy Movies}). Users can scroll vertically to browse across groups and horizontally to explore videos within each group. 

Organizing search results may also provide an opportunity to communicate to users that their search intent has been understood. 
For example, by understanding that the query \userquery{"sonic t"} targets an \textit{unavailable} video facet (cf. Fig. \ref{fig:4-ooc-lolomo}), we can provide specific evidence that increases users' trust with the product (e.g., \textit{"Fans of `Sonic the Hedgehog' have watched these"}) and additional ways to pivot and explore the catalog (e.g., "pills" relevant to the unavailable video such as \textit{Myths \& Legends} and \textit{Chases}).

\section{Approach and Technical Challenges}
We now describe, on a high level, our approach (and some involved technical challenges) to providing search recommendations and meaningfully organizing results to satisfy search intents in the entertainment context.

\textbf{Producing search recommendations that effectively satisfy search intents:} 
The general search context and the query itself put restrictions on which videos qualify as relevant recommendations. 
For example, query-agnostic personalized rankings or unpersonalized popular videos are not relevant enough in the search context and may be detrimental to user satisfaction. Hence, we apply recommender systems approaches such as collaborative filtering techniques and heavily rely on user behavior to learn relevant recommendations.  This approach has challenges as it can increase popularity bias and may lower relevance of results.

\textbf{Meaningfully organize and explain search results:} We organize results in the form of cohesive groups of videos, where each group has a distinct, descriptive label. Such organization has advantages of \textit{uniformity} (uniform across the Netflix Homepage and Search), \textit{familiarity} (users have been using such organization on the Homepage for a while), and \textit{versatility} (same layout can work across various intents and query facets). There are some technical challenges though, for example, how to group videos, how to label the groups, how to rank the groups, and how to rank videos in each of the groups. Additionally, to achieve optimal results organization, we need to consider (i) \textit{query specificity} (e.g. broad/narrow groups for under-specified/fully-specified queries), (ii) \textit{search intents} (e.g., where to place a video in the organization scheme based on whether the user may be \textit{fetching} or \textit{exploring}), and (iii) \textit{device idiosyncrasies} (e.g., higher efficiency of navigation on TVs).

\textbf{Handling of short queries:} Because typing queries on devices such as TVs is challenging, we offer \textit{\instantsearch{}} and produce results after every query keystroke. This leads to many short and ambiguous queries \cite{lamkhede_sigir_2019} for which it is challenging to accurately infer query facets (e.g., inferring that \userquery{"sonic t"} targets the unavailable video \movietv{\textit{Sonic the Hedgehog}}{https://www.imdb.com/title/tt3794354/}). Similar to above, in order to overcome this challenge, we rely on which types of results users interact with. Properly predicting such facets is crucial to optimize and organize results for different intents.

\textbf{Low-latency \textit{Instant Search}:} While offering \textit{\instantsearch{}} has clear user benefits, it also creates challenges in organizing results on-the-fly due to latency considerations. Our approach also needs to be scalable to searches from more than 200 million users globally.

\textbf{Putting things together:} Our implementation works as follows. For every search request, we retrieve search matches and search recommendations. These are blended together and ranked by a machine learned model that estimates relevance of each result for the given search context. The top results are then organized into possible cohesive and coherent groups using editorially-defined constraints. These groups are then ranked with an approach that is similar to the Homepage construction mechanism \cite{aro_blog_post}. The construction ensures removing duplicate group-result assignments and improving diversity of groups. The backend system communicates these groups (each group has a header and ordered results) to the user device where they are rendered.

\section{Conclusions and Future Work}
Providing recommendations in the search context and effectively organizing results adds meaningful value for users of entertainment services such as Netflix. There are many interesting challenges that we need to address moving forward, including: (i) improved query facet detection, (ii) learning the ranking and organization models in an end-to-end fashion, (iii) creating layouts containing heterogeneous user interface elements and models to showcase different types of results.

\begin{acks}
We are very thankful to our collaborators Vickie Zhang, Priya Kothari, Weidong Zhang, Ivan Provolav, Mike Galassi, Sudeep Das, Chris Steger, and Aish Fenton as well as internal reviewers Yves Raimond and Justin Basilico.
\end{acks}

\section*{Bios}
\textbf{Sudarshan Lamkhede} is a Senior Research Engineer at Netflix Research where he leads the applied research for Search algorithms. Prior to Netflix, he led research / engineering for various Web Search, Page Optimization and Personalization algorithms at Yahoo! Research. He holds a Master's in Computer Science from University of Buffalo and a Bachelor's in Computer Engineering from University of Pune.

\textbf{Christoph Kofler} is a Senior Product Manager at Netflix, Inc. where he is responsible for the innovation of Search and Personalization algorithms. Previously, he held positions at Bloomberg L.P., Google Research, and Microsoft Research. Kofler received his Ph.D. degree from Delft University of Technology, Netherlands and his Master's degree from Klagenfurt University, Austria (both in Computer Science). For the work carried out during his Ph.D. program, he received the Google Doctoral Fellowship and the SIGMM Award for Outstanding Ph.D. Thesis.

\bibliographystyle{ACM-Reference-Format}
\bibliography{recsys2021}

\end{document}